\documentclass[shortnote,twocolumn]{jpsj2}
\usepackage{graphicx}

\def\runtitle{Disorder and Interactions on a 1D Chain}
\def\runauthor{Jonathan M \textsc{Carter} and Angus \textsc{MacKinnon}}

\title{Disorder and Interactions on a 1D Chain}

\author{%
Jonathan M \textsc{Carter}$^{1}$\thanks{jm.carter@imperial.ac.uk.} and
Angus \textsc{MacKinnon}$^{1,2}$\thanks{a.mackinnon@imperial.ac.uk.}
}

\inst{%
$^1$ Blackett Laboratory, Imperial College London, London SW7 2BW, UK.\\
$^2$ Cavendish Laboratory, Madingley Road, Cambridge CB3 OHE, UK.
}

\recdate{\today}

\abst
{
}

\kword{Anderson localization, disorder, interactions, 1D}

\begin{document}
\sloppy
\maketitle

\section{Introduction}
It is well established that in the presence of disorder electron
wavefunctions can become localized. Considerable numerical work has been
carried out for non-interacting systems with results reaching a
reasonable consensus: theory and experiment are in general qualitative
agreement. However, in 3D the determined value of the universal critical
exponent is markedly larger than the empirically measured value
\cite{rf:1a}. This seems to suggest that an essential factor is missing
from calculations: the obvious candidate is the electron-electron
interaction. Furthermore, some have claimed to observe a Metal-Insulator
transition in 2D contrary to the widely accepted scaling theory of
Anderson localization \cite{rf:1b}. This is often accredited to the
effect of interactions. Hence during the last 10 years attention has
been switching to this more difficult case. The central problem is that
the model becomes a many-body system and so the Hilbert space grows
quickly with system size. This renders an exact numerical calculation
far beyond computational capabilities. Nevertheless, several studies
have been accomplished, these suggest inclusion of interactions may
yield non-trivial behaviour. 

Shepelyansky \cite{rf:1} performed calculations on two interacting
particles. In 1D, interactions caused a large enhancement of
localization length. Other work showed that in 2D the effect is possibly
stronger leading to delocalization \cite{rf:2}. However, some caution is
required as the method fails to reproduce known non-interacting results
when interactions are switched off. 

The most successful method for treating the finite density problem is
the Density Matrix Renormalization Group (DMRG) approach \cite{rf:4}.
This works by performing a direct diagonalization but reducing the
Hilbert space by systematically discarding basis states that do not
contribute significantly to the ground state. Applying this method to
the Anderson interacting model (defined in equation \ref{eq1}), a
delocalized regime was found for attractive interactions \cite{rf:4a}.
In more recent papers by the same authors, it was noted that interesting
physics is washed out in the averaging process. Charge reorganisations
can be seen as electrons on a chain shift from the Mott insulator limit
to (strong interactions) to the Anderson insulator limit (strong
disorder) \cite{rf:4b}. Extensions of DMRG to 2D have encountered
difficulties. 

We have developed a new method incorporating some of the ideas of DMRG
and the transfer matrix method successfully used in the non-interacting
case \cite{rf:3}. Section 2 describes the method and section 3 discusses
current progress. 

\section{A New Method}
The method was developed using an interacting Anderson model:

\begin{eqnarray}
\label{eq1}
\hat{H}&=&\sum_i\epsilon_i \hat{c}_i^{\dagger}\hat{c}_i+V\sum_i(\hat{c}_i^{\dagger}\hat{c}_{i+1}+h.c.)\nonumber \\
&+&U\sum_i(\hat{c}_i^{\dagger}\hat{c}_i)(\hat{c}_{i+1}^{\dagger}\hat{c}_{i+1})-\mu \sum_i \hat{c}_i^{\dagger}\hat{c}_i
\end{eqnarray}
where $\hat{c}^{\dagger}_i$ is the creation operator for site $i$. Site
energies $\epsilon_i$ are chosen from a box distribution of width $W$.
The hopping parameter is $V$ and interaction strength given by $U$. The
calculation is done in the grand canonical scheme with chemical
potential $\mu$. 

The chain is grown by repeatedly adding sites to both ends. Given a set
of eigenstates for a chain of length $L$, it is possible to form a basis
for a chain of length $L+2$ by considering the possible occupancies of
the two new end sites. An eigenstate $|\phi^L\rangle$ generates four
basis states $|0\phi^L0\rangle$, $|1\phi^L0\rangle$, $|0\phi^L1\rangle$
and $|1\phi^L1\rangle$ where the $0$ and $1$ signify the occupancy of
the two new end sites. Therefore, a general state of length $L+2$ and
$N$ electrons can be expanded as a linear combination of such basis
states: 

\begin{eqnarray}
|\psi^{L+2,N}\rangle&=&\sum_i a_i |0\phi_i^{L,N}0\rangle\nonumber\\
&+&\sum_j (b_j |1\phi_j^{L,N-1}0\rangle+c_j |0\phi_j^{L,N-1}1\rangle)\nonumber\\
&+&\sum_k d_k |1\phi_k^{L,N-2}1\rangle
\end{eqnarray}

Note that particle number is a good quantum number so it is only
necessary to expand in basis states where the total occupancy of
$|\phi^L\rangle$ and the new sites is $N$. A Hamiltonian matrix may then
be calculated for each particle number. This is done in the $L+2$ basis
with the $|\phi^{L}\rangle$ eigenvalue on the diagonal plus elements due
to addition of the end sites. Note that the eigenvectors of the system
of length $L+2$ are represented in the basis of the eigenvectors of the
system of length $L$ as expanded by the new sites. This implies that the
calculation of the effective Hamiltonian for $L+2$ sites is actually
performed in the expanded basis of the eigenvectors for $L+2$. 

After calculating all the elements, the matrix is diagonalized with the
eigenstates forming the basis states for the next iteration. 

The approximation introduced to enable the calculation to fall within
computational limits is to systematically remove basis states after each
iteration. The simple approach is to retain the low energy states, as
these are the only states likely to contribute to the next iteration
ground state. 

Most of the traditional methods for extracting the degree of
localization do not carry across into the many--body case. Fortunately,
phase sensitivity to boundary conditions does not suffer from this
problem. This must be implemented as a perturbation because the
calculation uses open boundary conditions. It turns out that this is
equivalent to calculating off--diagonal elements of the reduced density
matrix of the ground state, where the interior part of eigenstates is
summed over. This is intuitively obvious because only the ends contain
the relevant physics and we are interested in the probability of an
electron entering one end and an electron appearing at the other. This
procedure motivates simultaneously adding sites to both ends of the
chain. 

\section{Preliminary Results}
Initially, even in the non--interacting limit exponential decay was not
observed in the middle of the band. However, it was found that this was
an artifact of the particular criterium used for removing basis states.
At first a set number of states was retained after each iteration, but
this failed to reproduce exponentially localized behaviour even without
interactions.
It was then found this could be fixed by using a constant energy cutoff above
the ground state and thereby allowing the number of states retained at
each stage to fluctuate.  However, the method still shows a significant
dependence on the size of the energy cutoff and the number of states below
the cutoff does not rise with system size as might be expected.  We belive
this to be due to the absence of the downward  ``pressure'' of the eliminated
states.  We are currently investigating methods for compensating for this
effect.

\begin{figure}[h]
\begin{center}
\includegraphics[width=\linewidth]{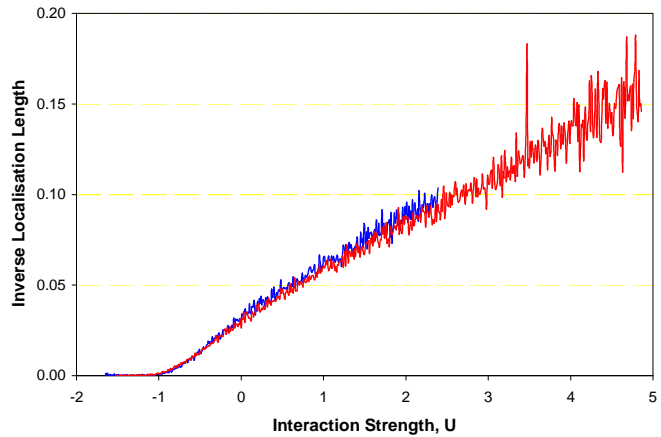}
\caption{Dependence of inverse localization length on interaction strength $U$ in the
middle of the band. Disorder $W=2$, $V=0$, $\mu=0$. The average number
of retained states at each step is $60$ -- continuous line, $80$ -- dotted
line.}
\label{fig1} 
\end{center}
\end{figure}
With these cautions, preliminary results have been obtained for the
middle of the band (half filling). Schmitteckert\cite{rf:4b} \textit{et
al} expected that repulsive interactions will enhance the effect of
disorder, whereas attractive interactions reduce it. Our preliminary
results are in agreement with this (figure \ref{fig1}). Furthermore, a
delocalized region for $U<-1$ was expected and observed using the DMRG
method. The results presented here are consistent with this prediction,
because our localization length diverges as $U$ approaches $-1$. 

\section{Conclusions}
Further refinement of the Hilbert space reduction procedure is required
before this method can be deemed reliable. Already extensions have been
made for the Hubbard model and a double chain model. It is worth noting
that the latter does not seem to suffer from the well--known sign
problem\cite{rf:5}. It is hoped that this method can be eventually applied to the
system with a finite cross section. 



\begin{thebibliography}{99}
\bibitem{rf:1a} K. Slevin and Tomi Ohtsuki: Phys. Rev. Lett {\bf 82} (1999)
382.
\bibitem{rf:1b} S. V. Kravchenko, G. V. Kravchenko and J. E. Furneaux: Phys. Rev. {\bf B50} (1994)
8039.
\bibitem{rf:1} D. L. Shepelyansky: Phys. Rev. Lett {\bf 73} (1994) 2607.
\bibitem{rf:2} M. Ortu\~no and E. Cuevas: Europhys. Lett. {\bf 46} (1999)
224.
\bibitem{rf:3} A. MacKinnon and B. Kramer: Phys. Rev. Lett. {\bf 47} (1981)
1546, Z. Phys. {\bf B51} (1983) 1.
\bibitem{rf:4} S. R. White: Phys. Rev. Lett. {\bf 69} (1992) 2863, Phys. Rev. {\bf B48} (1993)
10345.
\bibitem{rf:4a} P. Schmitteckert, R. A. Jalabert, D. Weinmann and J-L. Pichard: Phys. Rev. Lett {\bf 81} (1998)
2308.
\bibitem{rf:4b} P. Schmitteckert, T. Schulze, C. Schuster, P. Schwab and U. Eckern: Phys. Rev. Lett {\bf 80} (1998)
560.
\bibitem{rf:5} K.E.~Schmidt and M.H.~Kalos: in {\it Monte Carlo Methods in Statistical Physics II\/}, ed. K. Binder (Springer,
1984).
\end{thebibliography}
\end{document}